\documentclass[english,aps,amssymb,prb,twocolumn,showpacs]{revtex4}
\pdfoutput=1
\usepackage[T1]{fontenc}
\usepackage[latin9]{inputenc}
\usepackage{amsmath}
\usepackage{graphicx}
\usepackage{amssymb}
\usepackage{esint}

\makeatletter
\@ifundefined{textcolor}{}
{
 \definecolor{WHITE}{gray}{1}
 \definecolor{RED}{rgb}{1,0,0}
 \definecolor{GREEN}{rgb}{0,1,0}
 \definecolor{BLUE}{rgb}{0,0,1}
 \definecolor{CYAN}{cmyk}{1,0,0,0}
 \definecolor{MAGENTA}{cmyk}{0,1,0,0}
 \definecolor{YELLOW}{cmyk}{0,0,1,0}
 }

\renewcommand{\phi}{\varphi}
\renewcommand{\epsilon}{\varepsilon}

\renewcommand{\vec}[1]{{\bf #1}}

\usepackage{epsfig}
\makeatother

\usepackage{babel}

\begin{document}

\title {Majorana states in helical Shiba chains and ladders}

\author{Kim P\"oyh\"onen}
\author{Alex Weststr\"om}
\author{Joel R\"ontynen}
\author{Teemu Ojanen}
\email[Correspondence to ]{teemuo@boojum.hut.fi}

\affiliation{Low Temperature Laboratory (OVLL), Aalto University, P.~O.~Box 15100,
FI-00076 AALTO, Finland }

\date{\today}
\begin{abstract}
Motivated by recent proposals to realize Majorana bound states in chains and arrays of magnetic atoms deposited on top of a superconductor, we study the topological properties of various chain structures, ladders and two-dimensional arrangements exhibiting magnetic helices. We show that magnetic domain walls where the chirality of a magnetic helix is inverted support two protected Majorana states giving rise to a tunneling conductance peak twice the height of a single Majorana state. The topological properties of coupled chains exhibit nontrivial behaviour as a function of the number of chains beyond the even-odd dichotomy expected from the simple $\mathbb{Z}_2$ nature of coupled Majorana states. In addition, it is possible that a ladder of two or more coupled chains exhibit Majorana edge states even when decoupled chains are trivial. We formulate a general criterion for the number of Majorana edge states in multichain ladders and discuss some experimental consequences of our findings.

\end{abstract}
\pacs{73.63.Nm,74.50.+r,74.78.Na,74.78.Fk}
\maketitle
\bigskip{}

\section{Introduction}

Last few years have witnessed a great growth of interest in topological superconductivity and Majorana fermions in these systems.\cite{qi}  Majorana bound states (MBS) arise as Bogoliubov quasiparticles of topological superconductors and are always created in pairs.  Two spatially separated MBS can accommodate one ordinary fermionic state in a highly non-local fashion, a property that could be employed in quantum information storing and processing.\cite{kitaev1} In addition, the non-abelian statistics\cite{read, ivanov} obeyed by MBS make them promising candidates for topological quantum computing.\cite{nayak, kitaev2, kitaev3,alicea2}      

Recent experiments in semiconducting nanowires exhibit intriguing signatures of Majorana states.\cite{mourik, das, rokhinson, deng}  Building on previous work by Fu and Kane,\cite{fu} the existence of MBS in these systems was predicted independently by Lutchyn\cite{lutchyn} and Oreg\cite{oreg} and collaborators. The effective low-energy description of the nanowire model is equivalent to Kitaev's toy model, a paradigmatic one-dimensional (1D) topological superconductor.\cite{kitaev1} From an experimental point of view, the identification of MBS in nanowire systems is greatly complicated by effects of disorder that are unavoidably present. As understood recently, characteristic signatures of MBS such as zero-bias tunnelling peaks and the fractional Josephson effect may have alternative explanations. Novel realizations of MBS that could circumvent some of the complications of existing proposals are under active study.        

\begin{figure}
\includegraphics[width=0.8\columnwidth]{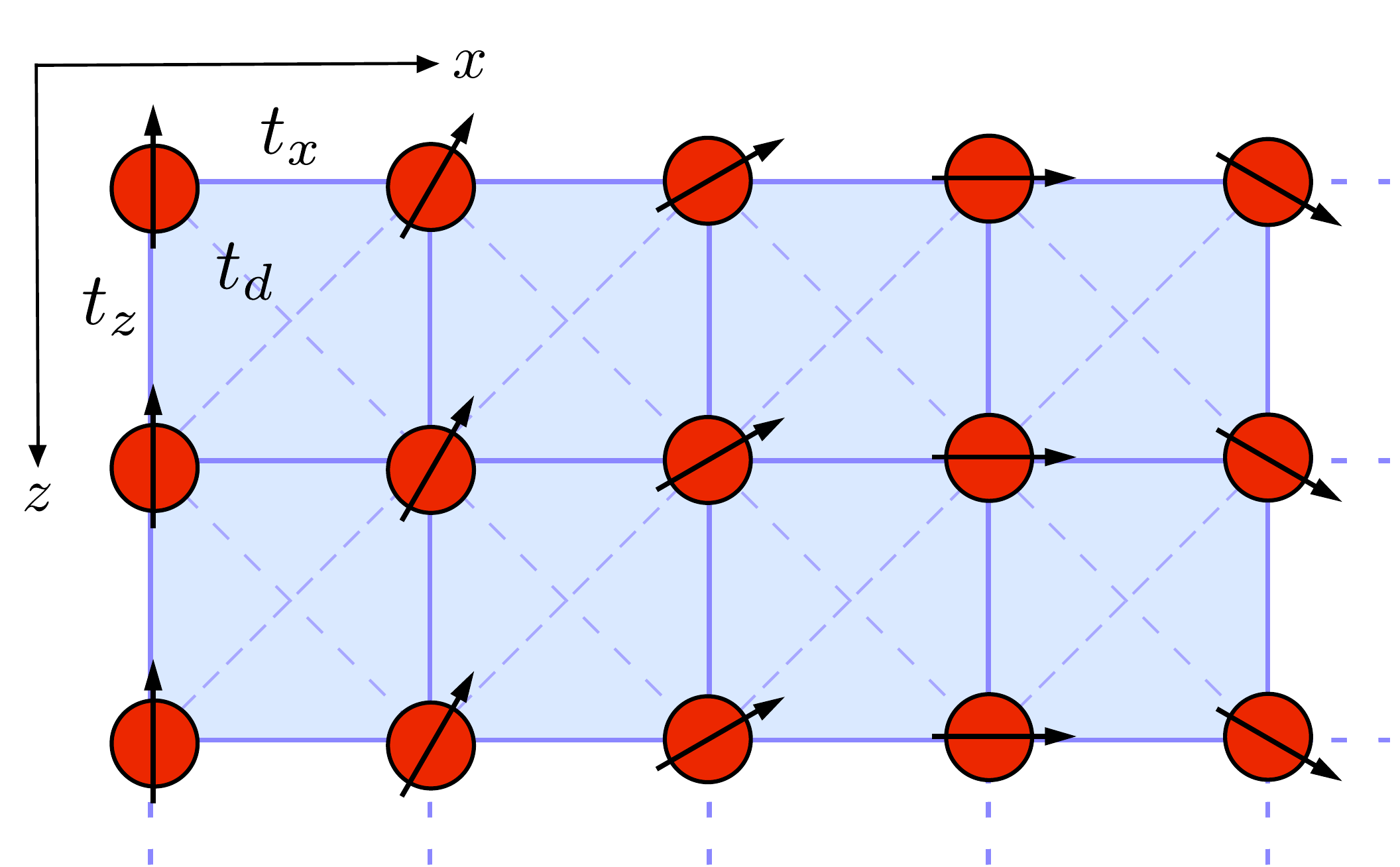}
\caption{ We study the properties of Shiba chains (corresponding to a single row) and ladders fabricated by depositing magnetic atoms on top of a superconductor by STM. The magnetic moments give rise to electronic impurity states that are coupled to each other. }\label{grid}
\end{figure}

A crucial ingredient of Kitaev's toy model complicating experimental realizations is the requirement for essentially spinless fermions. The necessary lifting of the spin degeneracy in the nanowire model is implemented by Rashba spin-orbit coupling in combination with a magnetic field. Various forms of magnetism also offer a possible route to topological superconductivity. As recently pointed out, helical magnetic order can effectively mimic Rashba coupling\cite{braunecker1,klinovaja2} which has motivated the pursuit for MBS in magnetic systems. Proposals for magnetic realizations of MBS can be roughly divided into two categories: systems based on engineering permanent magnets in the vicinity of nanowires\cite{klinovaja, ,kjaergaard, ojanen2} and systems arising naturally from microscopic interactions. The advantage of the systems in the first category is that they can be manipulated and controlled similarly to Rashba wires; however, they also share some of their weaknesses. Examples of systems in the second category are arrays of magnetic atoms deposited on top of a superconductor.\cite{choy,np,pientka2, vazifeh,klinovaja3,braun} Magnetic atoms can create Yu-Shiba-Rusinov impurity sites\cite{yu, shiba, rusinov, salkola} effectively forming a lattice. Coupled sites enable electron hopping from one site to another giving rise to an impurity band. The magnetic moments of the atoms lift the spin-degeneracy on the impurity site and the system may become topologically nontrivial, thus supporting MBS. The advantage of the impurity chain is that the sites can be deposited very accurately in the desired formation using STM, therefore eliminating an important source of disorder.\cite{np} In addition, impurity sites can be accessed individually and hence searching for signatures of MBS in the tunneling spectrum is flexible. Therefore these systems could potentially serve as an accurate testing ground for the generic properties of 1D topological superconductors.

Motivated by the model of an atomic chain with helical magnetic order, we study the topological properties and Majorana states of magnetic domain walls, quasi-1D structures and coupled chains. We begin by revisiting the properties of the 1D model\cite{np} and predict the existence of two protected zero-energy MBS in a magnetic domain wall where the chirality of the magnetic spiral is inverted. These MBS are protected by chiral symmetry and cannot hybridize when the magnetic texture is confined to the plane.  Therefore the tunnelling conductance to the domain-wall sites should exhibit a zero bias peak $2\times\frac{2e}{h}$ in the low-temperature limit, twice the height associated with a single MBS. After this we focus on ladders created by coupling 1D chains and discuss the topological properties of these systems.  In a simple physical picture developed below we explain how ladders can be thought of as a collection of independent transverse channels, a number of which can be in the nontrivial state. We provide a rule for the number of Majorana end states in ladders in Fig.~\ref{grid} and study effects of unideal features by numerical calculations. We point out that number of MBS does not follow from the simple even-odd variation of number of coupled chains in the ladder, as  expected from the $\mathbb{Z}_2$ character of interacting Majorana end states. We conclude by discussing some experimental consequences of our findings and summarizing our results.

\section{Majorana states in chains}

\begin{figure}
\includegraphics[width=0.8\columnwidth]{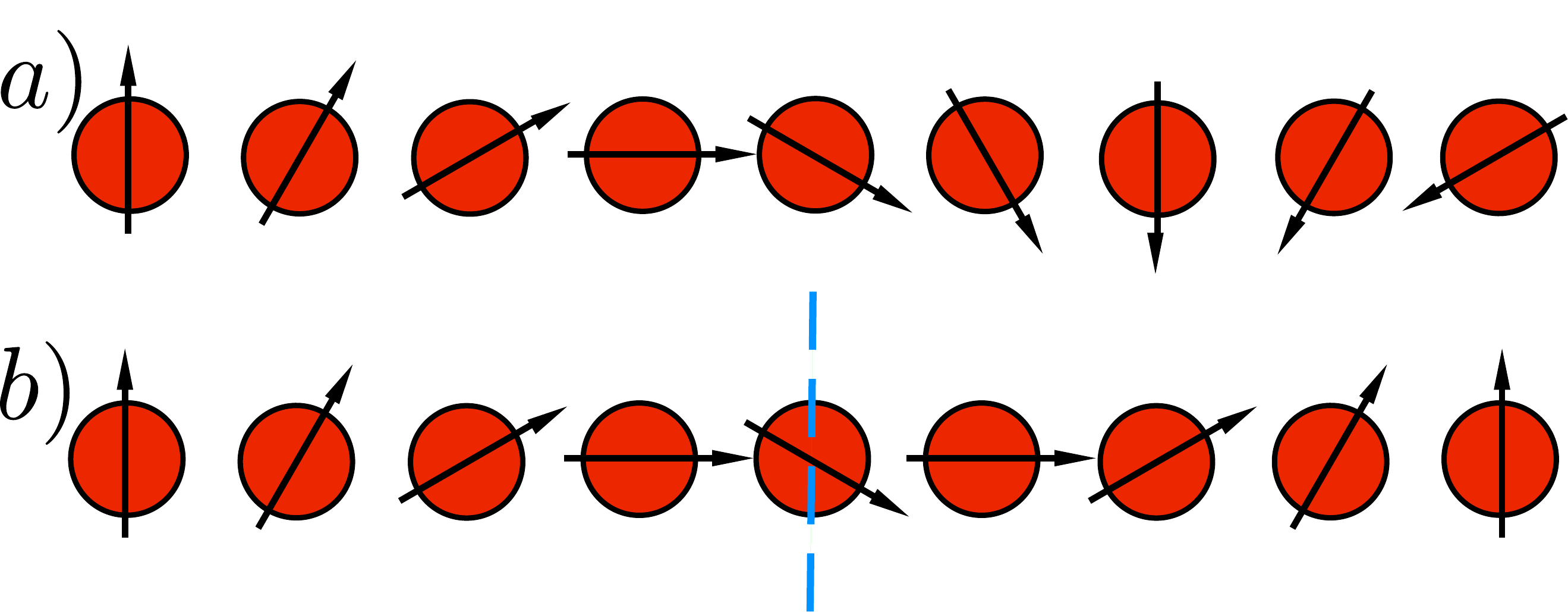}
\caption{a) Helical magnetic order in the chain may give rise to topologically nontrivial phase and Majorana states localized at both ends.  b) Domain wall separating two regions of opposite chirality supports two Majorana states that are degenerate for a planar texture. }\label{chains}
\end{figure}

\subsection{Topological properties of simple chains}
Here we revisit the 1D chain of helical magnetic impurity sites on a superconductor studied in detail in Ref.~\onlinecite{np}. A helical texture could arise from the RKKY interaction between localized moments mediated by electron-hole pairs\cite{braun, klinovaja3,vazifeh} but spin-orbit coupling arising from the surface effects may be important. Therefore we treat the pitch of the helix as a free parameter. For random impurity orientations the model has been studied in Ref.~\onlinecite{choy} where it was pointed out that the topological regime can also be achieved in this case. A microscopic derivation of an effective hopping model with similar low-energy properties was recently considered in Ref.~\onlinecite{pientka2}, the main difference being the long-range nature of hopping. In Sec.~\ref{dis} we discuss the relationship between these models.  

The Hamiltonian of a single horizontal chain in the ladder depicted in Fig.~\ref{grid} and shown in detail in Fig.~\ref{chains} a)  is described by
\begin{align}\label{h1d}
&H_{1d} = t_x\displaystyle\sum_{n\alpha}(f^\dag_{n\alpha}f_{n+1\alpha}+h.c.)-\mu\displaystyle\sum_{n\alpha}f^\dag_{n\alpha}f_{n\alpha}+\nonumber\\ 
&\displaystyle\sum_{n\alpha\beta}(\vec{B}_n\cdot\vec{\sigma})_{\alpha\beta}f^\dag_{n\alpha}f_{n\beta} + \Delta\displaystyle\sum_{n}(f^\dag_{n\uparrow}f_{n\downarrow}^\dag +h.c.),
\end{align}
where the first term arises from electron hopping with amplitude $t_x$, the second term fixes chemical potential $\mu$, an effective Zeeman field is given by $\vec{B}_n$ and the term proportional to $\Delta$ originates from superconducting pairing. The Pauli matrices acting on spin are denoted by $\sigma_i$. In the following we assume a helical magnetic texture in the lattice $\vec{B}_n = B_0\lbrace\sin(n\theta)\hat{\textbf{\textit{x}}}+\cos(n\theta)\hat{\textbf{\textit{z}}}\rbrace$ where $\theta$ determines the relative angle of the Zeeman field on two adjacent sites. We concentrate on planar textures $\vec{B}_n\cdot{\hat{\vec{y}}}=0$ unless otherwise stated.  Introducing the Nambu spinor $\Psi_i=(f_{i\uparrow}, f_{i\downarrow},f^{\dagger}_{i\downarrow},-f^{\dagger}_{i\uparrow} )^T$, we can write Eq.~(\ref{h1d}) in the Bogoliubov-de Gennes (BdG) form $H = \frac{1}{2}\sum_{ij}\Psi_i^\dag H_{ij}\Psi_j$ with
\begin{align} \label{tb}
H_{ij}= t_{x}\tau_z&(\delta_{i+1,j}+\delta_{i-1,j})+\nonumber\\
&+\delta_{ij}\left(-\mu \tau_z+\vec{B}_i\cdot\vec{\sigma} +\Delta\tau_x\right),
\end{align}
where matrices $\tau_i$ are Pauli matrices operating in the particle-hole space. By assuming that the spin rotation angle $\theta$ is a rational fraction of $2\pi$, it is possible to impose periodic boundary conditions for chains where the spin rotates an even number of times. The spectrum of the system and the Pfaffian invariant determining the topological phase was computed in Ref.~\onlinecite{np}. As expected, the signature changes of the Pfaffian coincide with the gap closing of the spectrum and the condition for the topologically nontrivial phase becomes 
\begin{align} \label{cond1}
\!\!\!\!\sqrt{\Delta^2+\!(|\mu|+2|t_x\alpha|)^2)}>\!B_0\!\!>\!\sqrt{\Delta^2+\!(|\mu|-2|t_x\alpha|)^2)}\!
\end{align} 
where $\alpha=\cos{(\theta/2)}$. \cite{np} While this result was derived for periodic systems, it is also relevant for systems with open boundary conditions since these exhibit MBS localized at the ends of the chain when the condition (\ref{cond1}) is satisfied. However, we would like to point out that the $\mathbb{Z}_2$ classification by the Pfaffian invariant does not provide a complete picture of topological properties. For a general magnetic texture (not necessarily planar), the BdG Hamiltonian (\ref{tb}) exhibits particle-hole symmetry $\{H_{ij},\tau_y\sigma_yK\}=0$, where $K$ denotes complex conjugation. However, in the studied case of planar helical magnetic order $H_{ij}$ also possesses chiral symmetry $\{H_{ij},\tau_y\sigma_y\}=0$ and thus the system actually belongs to the BDI class in the topological classification.\cite{schnyder} This fact has important consequences: chiral symmetry actually imposes a reality condition, so the effective low-energy description in  terms of a Kitaev-like model contains only real hopping amplitudes. This is also the reason why relaxing the requirement of planar texture leads to complex hopping terms in the low energy model, discussed in the context of a microscopic derivation of the Shiba-chain model.\cite{pientka2}  Furthermore, topological sectors of the BDI symmetry class are characterized by $\mathbb{Z}$-valued invariants in 1D. As a consequence, it is possible that domain walls and phase boundaries of model (\ref{tb}) support multiple protected MBS as shown below.  An approximate chiral symmetry may have important implications since a small chiral symmetry-breaking Zeeman  field component lifts the degeneracy of multiple MBS but does not annihilate them. The topological implications of chiral symmetry in 1D systems have been studied previously in  superconducting wires\cite{diez,ojanen1,stic} and normal systems.\cite{vayrynen}   
 
 \begin{figure}
\includegraphics[width=0.5\columnwidth]{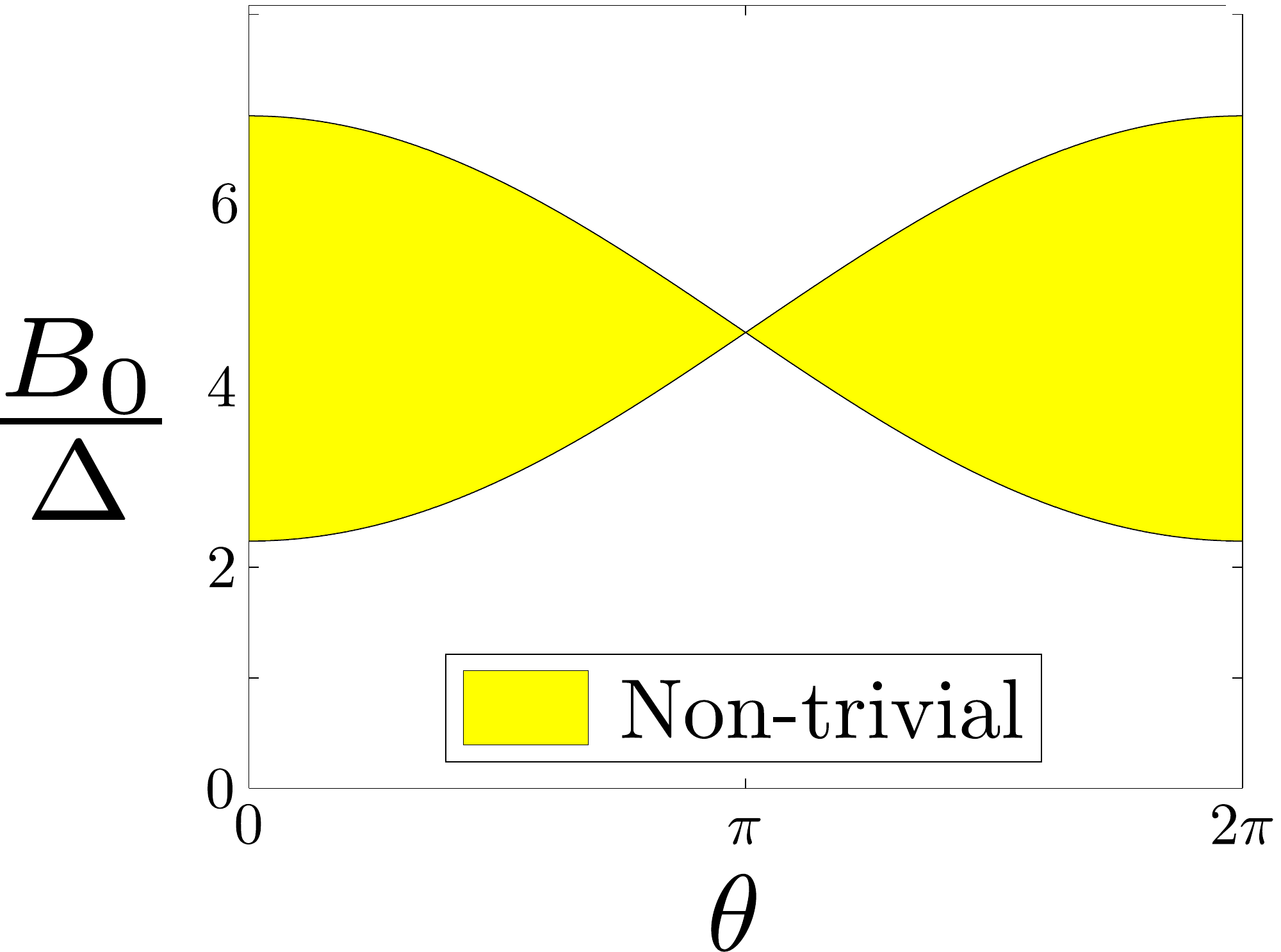}
\caption{ Topological phase diagram for a planar texture as a function of the pitch angle $\theta$ and Zeeman energy for $t_x=\Delta$, $\mu=4\Delta$. The energy gap vanishes near $\theta=0$ and $\theta=2\pi$, therefore the robust nontrivial states are located in the interior region away from the boundaries.}\label{cha1}
\end{figure}

\subsection{Majorana states at magnetic domain walls } \label{dw}

Having established the basic properties of simple helical chains, we now concentrate on a situation where the helix changes its chirality at a domain wall. This situation is depicted in Fig.~\ref{chains} b), where the clockwise rotation of the Zeeman field changes to counterclockwise rotation implemented by changing $\theta\to -\theta$ at the domain wall. The condition (\ref{cond1}) based on $\mathbb{Z}_2$ classification does not distinguish two nontrivial states corresponding to opposite chiralities. An appropriate $\mathbb{Z}$-valued winding number invariant of the BDI class can do this. Since a rotating magnetic field is closely related to the Rashba spin-orbit coupling, the magnetic domain wall is analogous to inverting the sign of the Rashba coupling.\cite{ojanen1} As pointed out  in Ref.~\onlinecite{ojanen1}, the topological phase diagram has three different sectors where the nontrivial phase is split into two separate phases. In analogy to the Rashba wire model, we also expect to find two degenerate MBS located at the boundary between two chiralities.

Numerical solution of the spectrum of Hamiltonian (\ref{tb}) with a magnetic domain wall $\theta\to-\theta$  confirms our qualitative expectations. In Fig.~\ref{dw} we have plotted the low-lying energy bands as a function of the Zeeman energy. For a finite range of $B_0$ there exist four mid-gap MBS separated from the other states by a gap. Two of the states are located at each end of the chain and two are located at the domain wall. Analogously to the case studied in Ref.~\onlinecite{ojanen1}, the magnetic domain wall separates two distinct topological sectors and therefore support two MBS.  

Since for a range of values of $B_0$ the magnetic domain wall plays the role of a phase boundary of two sectors, the two MBS located at the domain wall are expected to be insensitive to the details of the domain wall. A specific shape and size of the domain wall is not relevant for the existence the MBS as long as it interpolates between opposite chiralities. In particular, it is not necessary that $B_0$ and $|\theta|$ have the same asymptotic values on both sides. The two MBS cannot hybridize to finite-energy states as long as chiral symmetry remains intact and the overlap between the domain-wall states and the MBS located at the chain ends vanishes. Possible chiral-symmetry breaking perturbations include Zeeman field $\vec{B}_y$  in the $y$ direction and the superconducting phase gradient arising from a finite supercurrent. Surprisingly our calculations indicate that weak $\vec{B}_y$ is not sufficient in lifting the degeneracy of the domain-wall states.     

\begin{figure}
\includegraphics[width=0.95\columnwidth]{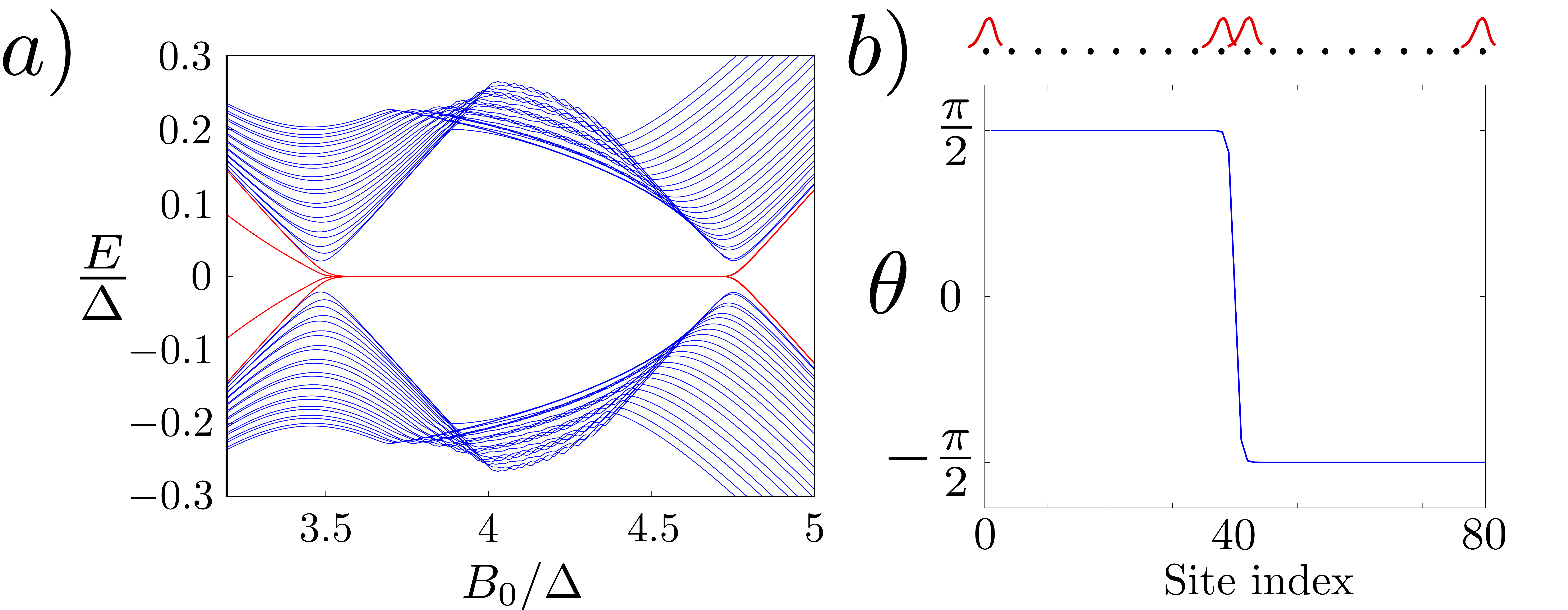}
\caption{a) Spectrum of a finite chain with 80 lattice sites with a domain wall in the middle corresponding to parameters $t_x=\Delta$, $\mu=4\Delta$. At critical Zeeman energy the system undergoes a topological phase transition and exhibits four zero energy Majorana bound states. Figure b) shows the domain-wall profile and indicates the locations of the four Majorana states corresponding to the zero-energy states in a). }\label{dw}
\end{figure}

The physical origin of the domain wall and the helix structure in general is an interesting question itself. Helical order may arise from the RKKY interaction between localized moments depending on microscopic details as well as the distance between the lattice sites. In addition to that there are surface effects that may favour one chirality over the other\cite{menzel}, making the detailed form of the structure difficult to predict. However, if the two opposite helical configurations are nearly degenerate, domain walls are likely to be present. Since atoms can be placed in arbitrary structures using STM, it could be possible to engineer domain walls artificially. Some implications of the domain-wall states to experimentally observable signatures are discussed in Sec.~\ref{dis}.  

\section{ Coupled chains}

The previous discussion of single chains provides sufficient ingredients to understand the properties of ladders constructed by coupling single chains. We concentrate mostly on systems similar to that depicted in Fig.~\ref{grid}, where the Zeeman field rotates along the longitudinal direction while different horizontal chains are coupled through the transverse hopping $t_z$ and diagonal hopping $t_d$. This geometry and magnetic texture is obviously just one among many possibilities, but interestingly the structure of a double chain of Fe atoms observed in Ref.~\onlinecite{menzel} is arranged in this type of configuration. Complementing the analytical treatment of identical horizontal chains, we study numerically more general effects of disorder in helical alignment as well as more complicated textures. 

\begin{figure}
\includegraphics[width=0.95\columnwidth]{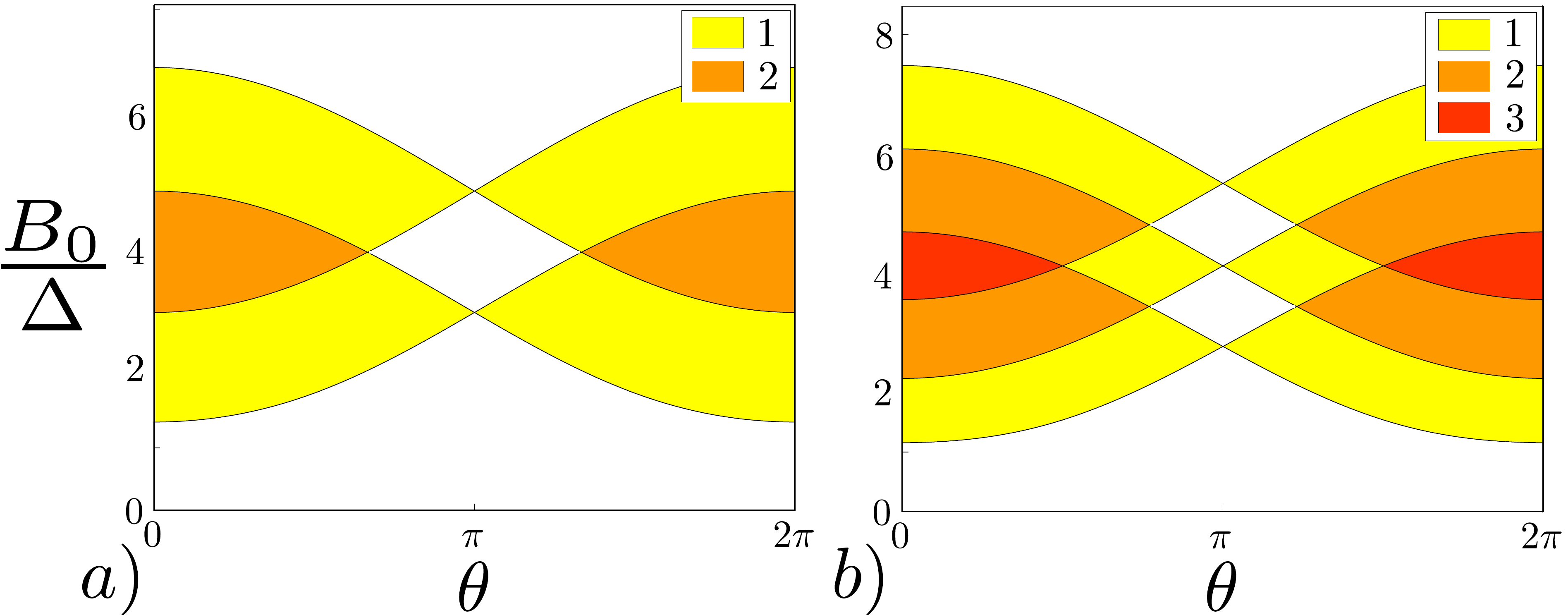}
\caption{a) Number of Majorana end states in a ladder of two coupled chains $N_z=2$ as a function of tilt angle and Zeeman field corresponding to parameters $t_x=\Delta$, $t_z=\Delta$, $t_d=0$, $\mu=4\Delta$ and $\theta=\pi/2$. b) same as a) but for three coupled chains.  }\label{cha2}
\end{figure}
The Hamiltonian of ladders with $N_x$  columns and $N_z$ rows can be expressed similarly to Eq.~(\ref{tb}), with two additional hopping terms coupling the horizontal chains. Denoting the position of the atoms in the ladder by $(i,j)$, the hopping part becomes 
\begin{align} \label{tb2}
 \frac{1}{2}\sum_{ij}&\left(t_x\Psi_{(i+1,j)}^\dag\Psi_{(i,j)}+t_z\Psi_{(i,j+1)}^\dag\Psi_{(i,j)}+\right.\nonumber\\
&+\left.t_d\Psi_{(i+1,j+1)}^\dag\Psi_{(i,j)}+\mathrm{h.c.}\right),
\end{align}
while the local part remains unchanged. The spectrum and Majorana states can be solved numerically as previously. However, in the case of a homogeneous ladder where the magnetic texture depends only on the horizontal coordinate (as in Fig.~\ref{grid}) it is possible to make analytical progress. In that case it is convenient to regard the ladder as a 1D system with an additional transverse sublattice structure of $N_z$ components. The field operators at horizontal site $i$ should be generalized to accommodate the sublattice structure $\hat{\Psi}_i=(\Psi_i^1,\Psi_i^2\ldots \Psi_i^{N_z})^T$. The BdG Hamiltonian in the new basis then takes the effectively 1D form  $H = \frac{1}{2}\sum_{ij}\hat{\Psi}_i^\dag H_{ij}\hat{\Psi}_j$ with
\begin{align} \label{tb3}
H_{ij}= (t_{x}+t_d\vec{S}) &\tau_z(\delta_{i+1,j}+\delta_{i-1,j})+\nonumber\\
+\delta_{ij}&\left[(-\mu+t_z\vec{S} )\tau_z+\vec{B}_i\cdot\vec{\sigma} +\Delta\tau_x\right],
\end{align}
where $\vec{S}$ is $N_z\times N_z$ hopping matrix operating on the sublattice degree of freedom $\vec{S}_{nm}=\delta_{n,m+1}+\delta_{n+1,m}$ with $n=1\ldots N_z-1$. To further simplify Eq.~(\ref{tb3}) we note that the eigenvalue problem in sublattice space $\vec{S}\vec{x}_p=\lambda_p\vec{x}_p$ can be solved exactly, where $\lambda_p=2\cos{[\frac{p\pi}{(N_z+1)}]}$    and $[\vec{x}_p]_q=\sqrt{\frac{2}{n+1}}\sin{[\frac{pq\pi}{N_z+1}]}$. After a unitary transform in the sublattice space  the Hamiltonian (\ref{tb3}) decouples to $N_z$ independent 1D expressions 
\begin{align} \label{tb4}
H_{ij}^p= (t_{x}+t_d\lambda_p) &\tau_z(\delta_{i+1,j}+\delta_{i-1,j})+\nonumber\\
+\delta_{ij}&\left[(-\mu+t_z\lambda_p )\tau_z+\vec{B}_i\cdot\vec{\sigma} +\Delta\tau_x\right].
\end{align}   
Thus we have explicitly derived the result stating that \emph{ a ladder of $N_z$ coupled chains is effectively described by $N_z$ independent 1D chains} with renormalized chemical potentials   $\mu_p=\mu -2\,t_z \cos{[\frac{p\pi}{(N_z+1)}]}$ and hopping amplitudes $t_{xp}=t_x+2\,t_d\cos{[\frac{p\pi}{(N_z+1)}]}$  for $p=1\ldots N_z$.
\begin{figure}
\includegraphics[width=0.95\columnwidth]{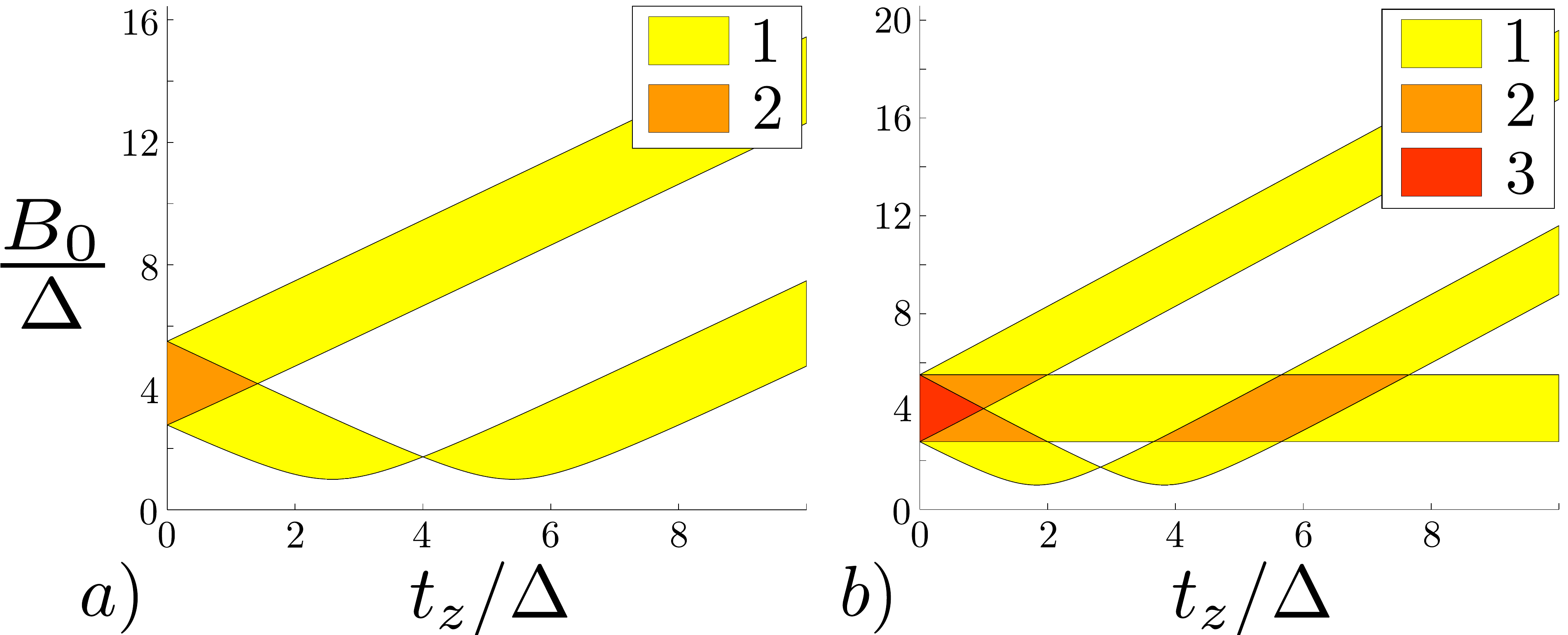}
\caption{ a) Colours indicate the number of MBS at each end (topological transverse channels) in a ladder of two coupled chains $N_z=2$ as a function vertical hopping. The plot parameters are $t_x=\Delta$, $t_d=0$, $\mu=4\Delta$ and $\theta=\pi/2$. b) Same as a) but for $N_z=3$. }\label{cha3}
\end{figure}

The result (\ref{tb4}) provides a general framework to discuss the topological properties of ladders. Since ladders decouple to independent 1D systems, their topological properties can be solved by considering transverse modes separately. The modes differ only by the values of the chemical potentials and hopping amplitudes and the nature of each mode is determined by Eq.~(\ref{cond1}), where $\mu$ should be replaced by $\mu_p$ and $t_x$ by $t_{xp}$ for each $p=1\ldots N_z$. Also, the renormalization effects from $t_z$ and $t_d$ are very similar from a topological point of view as is evident in Eq.~(\ref{cond1}) The maximum number of topological channels is obviously $N_z$, but any number between 1 and $N_z$ in general is possible depending on parameters. In Fig.~\ref{cha2} we have plotted topological phase diagrams for different numbers of transverse modes as a function of $\theta$. A comparison between Figs.~\ref{cha1} and \ref{cha2} immediately reveals how the parameter region of at least one Majorana end state is expanded in the case of multiple chains.  

\begin{figure}
\includegraphics[width=0.60\columnwidth]{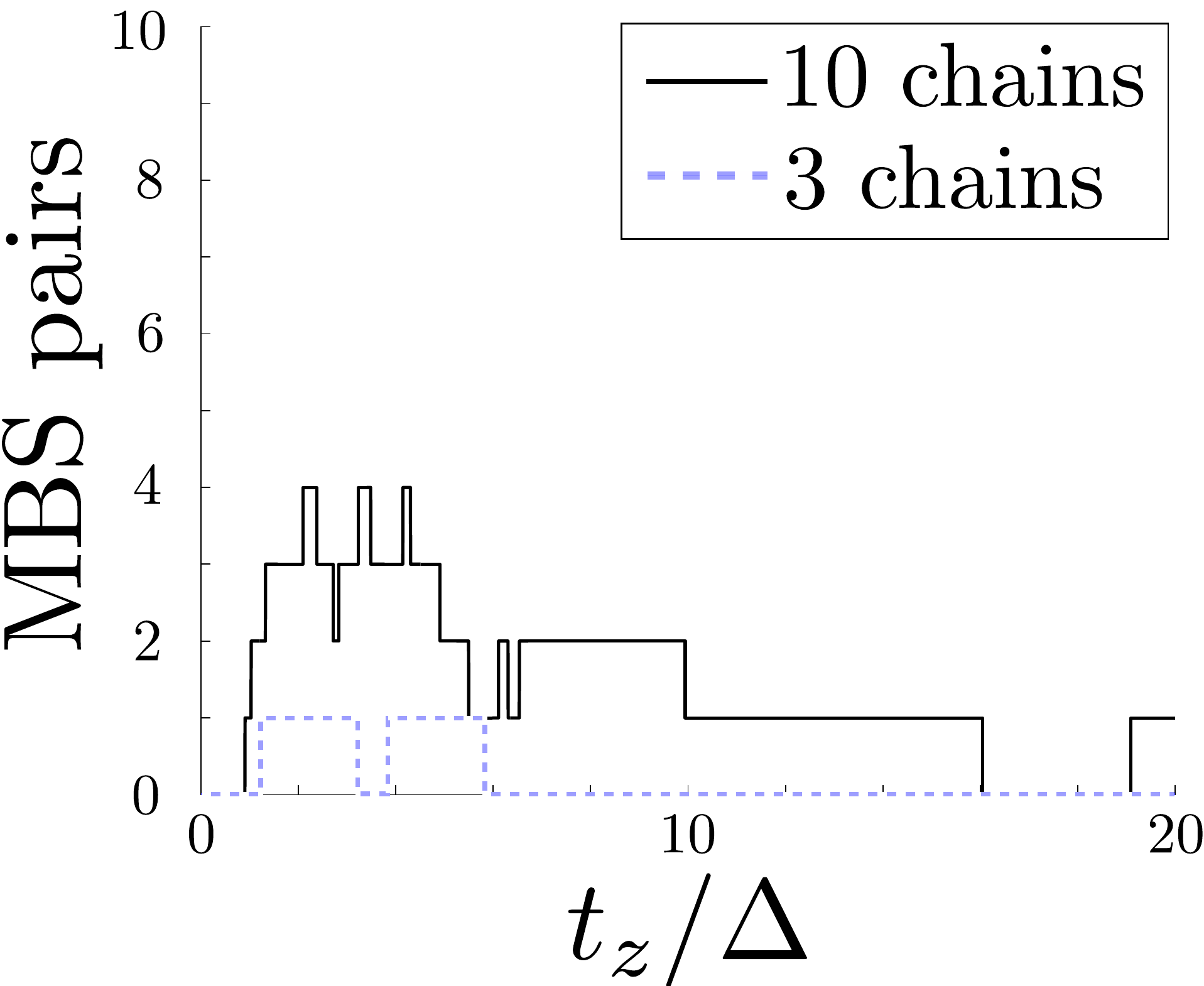}
\caption{ Number of topological channels (MBS pairs)  in the ladder as a function of transverse hopping amplitude in the case where decoupled chains are in the trivial phase. The plot parameters are $t_x=\Delta$, $t_d=0$, $\mu=5\Delta$, $B_0=2.1\Delta$ and $\theta=\pi/2$.   }\label{numb}
\end{figure}

In Fig.~\ref{cha3} we have plotted the number of topological channels as a function of the Zeeman energy and transverse hopping $t_z$. The rich nonmonotonic structure of the number of topological channels as a function of $t_z$  follows mathematically from the absolute value signs in Eq.~(\ref{cond1}). As illustrated in Fig.~(\ref{numb}), even when the individual chains are in the nontrivial phase, the number of topological channels and the MBS located at the ends of the chain do not follow the simple $\mathbb{Z}_2$ behaviour where the number of MBS is given by $N_z$ mod 2.  The only notable difference between even and odd $N_z$ in the studied model is that in the odd case the transverse spectrum $\lambda_p=\cos{[\frac{p\pi}{(N_z+1)}]}$ always contains solution $\lambda_p=0$. This implies that ladders with odd $N_z$  always contain at least one nontrivial channel if the channels are in the nontrivial phase when decoupled. Since vertical hopping expands the region of parameter space sampled by different transverse modes, ladders are more likely to exhibit MBS than single chains.  Interestingly, it is possible to achieve Majorana end states in some counterintuitive cases such as when two or more trivial chains are coupled by vertical hopping.

A direct numerical solution of the spectrum of the ladder confirms the analytical theory of the number of MBS. This analysis shows that the 1D chain is robust to site-to-site random perturbations of the magnetic moment angle. The topological modes, including those at the domain wall, were generally tolerant of small random fluctuations, in some cases remaining unaffected by an added random component taking values greater than $\frac{\theta}{3}$. As expected, larger values of $\theta$ allow for a larger absolute magnitude of the disorder. In the case of coupled chains, the situation is more complicated. In some cases, the system is as robust as in the 1D scenario while in some situations the number of topological channels can change more easily. Physically this behaviour is expected since the phase diagrams in Fig.~\ref{cha2} contain more boundaries that can be crossed by random fluctuations. Deep in a topological phase ladders are still very robust to $\theta$ disorder.  For example, for the parameters used in Fig.~\ref{cha2} b), letting $\theta=\pi/3$ and $B=4$ to place the system in the middle of the topological regime, the qualitative features of spectrum are unaffected by a uniform random onsite angle distribution with amplitude $\theta/2$.

We also studied ladders where the magnetic texture may rotate in the $z$ direction, keeping the horizontal pitch in the different chains identical. We discovered that ladders with identical horizontal rows and parameters placing them in the middle of a topological phase are generally insensitive to a linear variation of $\theta$ as a function of $z$ coordinate. 

\section{Discussion} \label{dis}

Above we have analyzed a model of magnetic atoms placed on a superconductor assuming that magnetic moments induce impurity states and self-assemble to a spiral magnetic state. While the studied topological properties and MBS follow from this relatively simple model, the applicability of the model to realistic systems is a relevant question. The existence of Shiba states is an experimentally established fact, but the helical order of the magnetic moments and the ability to tune the system to a topologically nontrivial parameter regime is not clear at the moment. There are theoretical arguments suggesting that the pitch of the helix will naturally adjust to the parameter regime supporting topological phase.\cite{vazifeh, klinovaja3, braun}  A comprehensive microscopic treatment requires a microscopic solution of the impurity problem and the formulation of a detailed theory of the interaction between magnetic impurity sites for which the surface effects may become important. An important step towards a theoretical understanding of the problem was taken in Ref.~\onlinecite{pientka2}, where an effective Kitaev-like model was derived from a microscopic Hamiltonian.  The qualitative difference between their model and Eq.~(\ref{h1d}) arises from the long-range nature of the hopping term in Ref.~\onlinecite{pientka2}. The complex phases in the hopping amplitudes in the presence of nonplanar helical texture also follow from Eq.~(\ref{h1d}) when it is projected to the low-energy bands.\cite{choy}

While a rigorous justification of the model is obviously beyond the scope of the present work, we will now discuss consequences of our predictions that could be tested in experiments.        
Shiba chains have the great advantage that MBS could be accessed directly through the tunnelling spectrum. Focusing an STM tip close to a specific impurity site can probe the local density of states. The existence of MBS give rise to a finite density of states within the superconducting gap and should show up as sub-gap peaks in differential conductance. The tunneling spectrum in the presence of multiple MBS was studied by Flensberg. \cite{flensberg} In the low-temperature limit $T\ll \Gamma$ where $\Gamma$ is the coupling between the STM tip and a MBS, the differential conductance should show a quantized peak $\frac{dI}{dV}=\frac{2e^2}{h}\frac{4\Gamma^2}{(eV)^2+4\Gamma^2}$.\cite{flensberg} At finite temperatures the peak height is reduced but the enhanced density-of-states should be observable. If the tip can couple to several noninteracting MBS for which couplings all satisfy $\Gamma_i\gg T$ the expected conductance is a simple sum over all MBS. Therefore, the two degenerate overlapping MBS at the magnetic domain wall discussed in Sec.~\ref{dw} should give rise to a $\frac{4e^2}{h}$ peak in the low-temperature limit. When chiral symmetry is broken, for example by driving supercurrent in the system, the two MBS acquire an energy splitting and move to finite energies.  Then the individual contributions to the tunneling conductance take the form $\frac{dI}{dV}=\frac{2e^2}{h}\frac{(2eV\Gamma_{1/2})^2}{((eV)^2-4t^2)^2+(2eV\Gamma_{1/2})^2}$, where $t$ is the energy splitting of the MBS.\cite{flensberg} The maximum peak height from the sum of these contributions remains $\frac{4e^2}{h}$ but the peak splits to two separate peaks achieving maxima at  $eV=\pm 2t$. The reason why enhancement from multiple MBS in the tunneling spectrum could be observed is that the chain may support multiple degenerate (or nearly degenerate) overlapping MBS making it possible for more than two MBS to satisfy $\Gamma_i\gg T$. The domain-wall MBS are protected by chiral symmetry even though they are overlapping in space. In the absence of extra symmetries, two nearby MBS eventually merge to the gap edge and the sub-gap tunneling peaks vanish.  

The enhancement of tunneling conductance from multiple Majorana edge states in ladders could be observed in a similar manner. Measuring the tunneling spectrum from the ends of the ladder should enable coupling to multiple MBS in the topological regime, therefore leading to strongly enhanced signal compared to a case of a single MBS. Studying structures with different numbers of coupled chains could reveal the predicted pattern of MBS. Since the parameter regime where the system exhibits at least one topological channel is enhanced in ladders, they could prove fruitful in observing Majorana states.       

\section{Conclusion}
In this work we studied the topological properties of a chain of magnetic impurity sites on a superconductor. Our main findings are the discovery of two Majorana bound states on magnetic domain walls where helical magnetic order inverts chirality and a systematic study of Majorana states in multichain ladders with spiral magnetic order. The number of Majorana bound states in ladders has a rich structure and exhibits counterintuitive features. The domain-wall Majorana states can be expected to show up as an enhanced peak in the tunneling spectrum. The predicted pattern of Majorana end states in ladders could also be observed by fabricating systems with different numbers of coupled chains and studying the tunneling spectrum.       

\acknowledgements
The authors would like to thank the Academy of Finland for support.

\end{document}